\documentclass[aps,prl,twocolumn,showpacs]{revtex4}  

\usepackage{graphicx}  
\usepackage{epstopdf}
\usepackage{dcolumn}   
\usepackage{bm}        
\usepackage{amsmath}
\usepackage{amssymb}   
\usepackage{array}
\usepackage[caption=false]{subfig}
\usepackage[bookmarks=false,linkcolor=blue,urlcolor=blue,colorlinks,citecolor=blue]{hyperref}
\usepackage{exscale,relsize}
\usepackage{bbm}

\def \w{{\rm w}}
\def \s{{\rm sc}}
\def \k{{k}}
\def \top{{\rm T}}
\def \mk{{-k}}
\def \bs#1{{\boldsymbol#1}}
\def \tx#1{{\rm#1}}
\def \mc#1{\mathcal{#1}}

\DeclareMathOperator{\Real}{Re}
\makeatletter

\begin{document}

\title{No-go theorem for a time-reversal invariant topological phase in noninteracting systems coupled to conventional superconductors}

\author{Arbel Haim$^1$, Erez Berg$^1$, Karsten Flensberg$^2$, and Yuval Oreg$^1$}
\affiliation{$^1$Department of Condensed Matter Physics$,$ Weizmann Institute of Science$,$ Rehovot$,$ 76100$,$ Israel\\
\mbox{$^2$Center for Quantum Devices, Niels Bohr Institute$,$ University of Copenhagen$,$ DK-2100 Copenhagen \O, Denmark}}
\date{\today}

\begin{abstract}
We prove that a system of non-interacting electrons proximity coupled to a conventional $s$-wave superconductor cannot realize a time reversal invariant topological phase. This is done by showing that for such a system, in either one or two dimensions, the topological invariant of the corresponding symmetry class (DIII) is always trivial. Our results suggest that the pursuit of Majorana bound states in time-reversal invariant systems should be aimed at interacting systems or at proximity to unconventional superconductors.
\end{abstract}

\maketitle

\emph{Introduction.---}Topological superconductors (TSCs) are characterized by a bulk superconducting gap and topologically protected subgap boundary excitation. The nature of these excitations depends on the dimensionality and the symmetries of the system~\cite{schnyder2008classification,kitaev2009periodic}. In symmetry class D~\cite{Altland1997}, with only particle-hole symmetry (PHS), and in one dimension (1d), these are the zero-energy Majorana bound (MBSs) states~\cite{kitaev2001unpaired,Alicea2012,Beenakker2013}.

Symmetry class DIII includes systems which in addition to PHS, possess a time-reversal symmetry (TRS) which squares to $-1$. This dictates a Kramers degeneracy of the single-particle spectrum. In the topological phase such a system would therefore host a \emph{pair} of zero-energy MBSs, related by time reversal~\cite{Qi2009time}. Indeed, such a time-reversal invariant topological superconductor (TRITOPS) can be thought of as two copies of a topological superconductor in class D. This is analogous to the topological insulator which is equivalent to two copies of a quantum Hall insulator.

Experimentally realizing the TRITOPS phase is a major outstanding challenge in the study of topological phases in condensed matter. To this date, however, attempts have been focused on realizing the class D TSC~\cite{mourik2012signatures,deng2012anomalous,Das2012zero,churchill2013superconductor,Finck2013anomalous,Nadj-Perge2014observation,Pawlak2015probing,Ruby2015end}. An important breakthrough in this context was the understanding that one can realize class D TSC using a combination of spin-orbit coupling and proximity to an $s$-wave superconductor, in a system of noninteracting electrons~\cite{fu2008superconducting,fu2009josephson,sau2010generic,Alicea2010Majorana,Lutchyn2010majorana,Oreg2010helical}.

In this Rapid Communication we prove that, unlike in class D, the topological phase of class DIII cannot be realized using proximity of a conventional $s$-wave superconductor to a system of noninteracting electrons. This was previously shown to be correct in two particular systems~\cite{Zhang2013time,Gaidamauskas2014majorana}. In this work we prove it for the most general system of noninteracting electrons, and the most general form of coupling to the superconductor.

Our result suggests that to realize the TRITOPS phase one should consider interactions between the electrons~\cite{Gaidamauskas2014majorana,Haim2014time,Klinovaja2014time,Klinovaja2014Kramers,Danon2015interaction,Pedder2015,Haim2016interaction}, or use proximity to unconventional superconductors~\cite{Wong2012majorana,Zhang2013time,Nakosai2013majorana}. One can also use \emph{two} $s$-wave superconductors with a phase difference between them which is tuned to $\pi$~\cite{Dahlhaus2010,Keselman2013inducing,Schrade2015proximity}. In 2d and 3d, intrinsic TRITOPS have been proposed~\cite{Fu2010,Nakosai2012topological,Deng2012majorana,Wang2014two,Brydon2014Odd,Scheurer2016mechanismNote,volovik2003universeNote}, which do not involve the proximity effect~\cite{GaplessTSC}.

\begin{figure}
\begin{tabular}{cc}
\rlap{\parbox[c]{0cm}{\vspace{-3.2cm}\footnotesize{(a)}}}
\hskip -1mm
\includegraphics[clip=true,trim =0mm 0mm 0mm 0mm,width=0.23\textwidth]{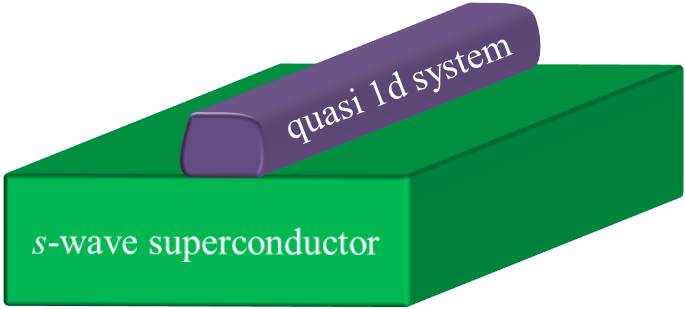} &
\rlap{\parbox[c]{0cm}{\vspace{-3.2cm}\footnotesize{(b)}}}
\hskip -1mm
\includegraphics[clip=true,trim =0mm 0mm 0mm 0mm,width=0.23\textwidth]{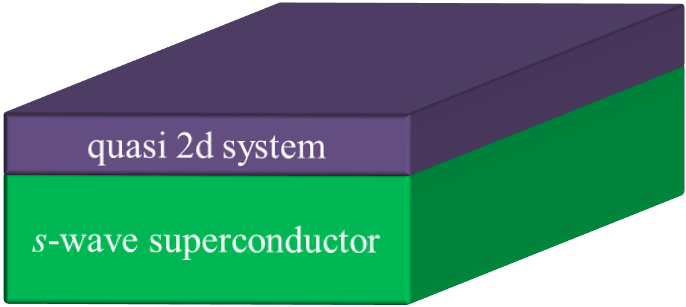}
\end{tabular}
\caption{The considered physical setups. (a) A quasi one-dimensional system (referred to as a wire) in proximity to a conventional $s$-wave superconductor. As long as there are no interactions between the electrons in the wire, the system will never be in the topological phase of class DIII. (b) This statement can be extended to the case of a quasi two-dimensional system.}\label{fig:setup}
\end{figure}

We start by writing the model, consisting of both the parent superconductor and the system as depicted in Fig.\hyperref[fig:setup]{~\ref{fig:setup}(a)}. We integrate out the superconductor's degrees of freedom and obtain the Green's function of the system alone. Next we construct the $\mathbb{Z}_2$ topological invariant for a general class-DIII system in 1d, and show that this invariant always takes its trivial value. We then extend this result to the case of a 2d proximitized system [see Fig.\hyperref[fig:setup]{~\ref{fig:setup}(b)}]. Finally, we generalize the proof to include also non translationally-invariant systems.

\emph{Model.---}We consider a quasi--1d system (hereafter referred to as a ``wire'') of noninteracting electrons, coupled to a bulk superconductor (SC). The Hamiltonian describing the combined system reads
\begin{equation}\label{eq:H}
\begin{split}
&H = H_\w + H_\s + H_\tx{c}, \phantom{\sum_k}\\
&H_\w = \sum_{\k} \psi_\k^\dag h^\tx{w}_\k \psi_\k,\\
&H_\s = \sum_\k \left[ \eta_\k^\dag h^\tx{sc}_\k \eta_\k + \frac{1}{2}(\eta_\k^\dag \Delta_\k \eta_\mk^{\dag\top} + \tx{h. c.})\right],\\
&H_\tx{c} = \sum_\k ( \eta_\k^\dag t_\k \psi_\k+\tx{h. c.} ),
\end{split}
\end{equation}
where $k$ is the momentum along the wire's axis. $H_\w$ and $H_\s$ are the Hamiltonians describing the wire and the SC, respectively, and $H_\tx{c}$ describes the coupling between them. For every $\k$, $\psi^\dag_\k$ and $\eta^\dag_\k$ are row vectors of fermionic creation operators of states in the wire and the superconductor, respectively. These states include all degrees of freedom within a unit cell including spin, transverse modes, sublattice sites, atomic orbitals etc. Correspondingly, $h^\tx{w}_\k$, $h^\tx{sc}_\k$, $\Delta_\k$, and $t_\k$ are matrices operating on these internal degrees of freedom. In writing Eq.~\eqref{eq:H} we have assumed that the interactions in the SC are adequately described within mean-field theory through the pairing potential matrix, $\Delta_\k$~\cite{ReverseProximityEffect}. Due to fermionic statistics one can, without loss of generality, take the pairing matrix to obey $\Delta^\top_\mk=-\Delta_\k$, where the superscript stands for the transpose of a matrix. In Eq.~(\ref{eq:H}), we have assumed that the system is translationally invariant; however, below we argue that our conclusions hold even in non-translationally invariant systems, e.g., in the presence of disorder.

Here, we consider systems which belong to symmetry class DIII. The Hamiltonian $H$ has TRS that squares to $-1$. The application of such a time-reversal operation most generally reads
\begin{equation}\label{eq:TR}
\begin{split}
&\mathbb{T}\psi_\k \mathbb{T}^{-1}=\mc{T}_\w \psi_\mk \hskip 5mm ; \hskip 5mm
\mathbb{T}\eta_\k \mathbb{T}^{-1}=\mc{T}_\s \eta_\mk \\
&\mathbb{T} i \mathbb{T}^{-1}=-i,
\end{split}
\end{equation}
where $\mc{T}_\w$ and $\mc{T}_\s$ are unitary matrices operating in the spaces of states in the wire and the superconductor, respectively, and which furthermore obey $\mc{T}^{\phantom{\ast}}_{\w(\s)}\mc{T}_{\w(\s)}^\ast=-1$. The last property is what distinguishes systems in class DIII from systems in class BDI~\cite{Dumitrescu2015majorana,Kotetes2015topological} in which TRS squares to $1$. Enforcing TRS on the system, $\mathbb{T}H\mathbb{T}^{-1}=H$, amounts to the following conditions
\begin{equation}\label{eq:TR_on_matrices}
\begin{split}
&\mc{T}_\w^\dag h^{\w\ast}_\mk \mc{T}_\w = h^\w_\k \hskip 5mm , \hskip 5mm
\mc{T}_\s^\dag h^{\s\ast}_\mk \mc{T}_\s = h^\s_\k, \\
&\mc{T}_\s^\dag t^\ast_\mk \mc{T}_\w = t_\k \hskip 7mm , \hskip 5mm
\mc{T}_\s^\dag \Delta^\ast_\mk \mc{T}_\s^\ast = \Delta_\k.
\end{split}
\end{equation}
The last equality, together with the property $\Delta_\mk^\top=-\Delta_\k$, guarantee that $\Delta_\k\mc{T}_\s$ is an Hermitian matrix.

In this work, we focus on the case where the pairing potential of the parent superconductor satisfies that $\Delta_\k\mc{T}_\s$ is a positive semi-definite (PSD) matrix~\cite{CmplxNumTimesPSD}. Namely $\langle u| \Delta_\k \mc{T}_\s |u\rangle \ge 0$ for all vectors $|u\rangle$, and all momenta $k$. In particular, this includes for example the case of a \emph{conventional s-wave superconductor}, in which the order parameter has a uniform phase on all the bands (and no interband pairing). Note also that this condition excludes both the case considered in Ref.~\cite{Zhang2013time}, where the superconductor has an $s_{\pm}$ order parameter with a relative $\pi$ phase between different bands, and the case of Refs.~\cite{Dahlhaus2010,Keselman2013inducing}, where there are two superconducting leads that form a $\pi$ junction.

The simplest example of a time reversal invariant superconductor with a positive semi-definite $\Delta_k \mathcal{T}_{\s}$ is a single band $s$-wave superconductor, whose pairing potential is
\begin{equation}\label{eq:H_sc_exp}
(\Delta_\k)_{\bs{k}^{\phantom{\prime}}_\tx{T},s;\bs k^\prime_\tx{T},s'} = \Delta_0 i\sigma^y_{ss'} \delta_{\bs{k}^{\phantom{\prime}}_\tx{T},-\bs k^\prime_\tx{T}} ,
\end{equation}
where $\{\sigma^\alpha\}_{\alpha=x,y,z}$ is the set of pauli matrices operating in spin space, $\bs{k}^{\phantom{\prime}}_\tx{T}$ labels the transverse momenta of states in the superconductor, and $\Delta_0$ is a number which we can choose to be real and positive. The time-reversal matrix is given in this example by $(\mc{T}_\s)_{\bs{k}^{\phantom{\prime}}_\tx{T},s;\bs k'_\tx{T},s'}=-i\sigma^y_{ss'} \delta_{\bs{k}^{\phantom{\prime}}_\tx{T},-\bs k'_\tx{T}}$, which indeed results in $\Delta_\k \mc{T}_\s=\Delta_0$ being a PSD matrix. In what follows we will not limit ourselves to the example of Eq.~\eqref{eq:H_sc_exp}, but rather consider the most general matrix $\Delta_\k$ for which $\Delta_\k\mc{T}_\s$ is PSD~\cite{SOCinSC}. In particular, the superconductor can have multiple bands.

Below we prove that as long as $\Delta_\k\mc{T}_\s$ is PSD, the wire is in the topologically-trivial phase. We do this in two steps. First, we show that upon integrating out the SC, the anomalous part of the zero-frequency self energy is also PSD. Second, we show that, as a result, the $\mathbb{Z}_2$ topological invariant always assumes its trivial value.

\emph{Integrating out the superconductor.---}We wish to obtain the Green's function describing the wire, where the superconducting proximity effect is expressed by an anomalous self-energy term. We start by writing the Hamiltonian in a BdG form
\begin{equation}\label{eq:H_BdG}
H=\frac{1}{2}\sum_\k \Psi_\k^\dag \begin{pmatrix} \mc{H}^\w_\k & V^\dag_\k \\ V_\k & \mc{H}^\s_\k \end{pmatrix} \Psi_\k,
\end{equation}
using the Nambu spinor $\Psi^\dag_\k =(\psi^\dag_\k , \psi^\top_\mk \mc{T}_\w , \eta^\dag_\k , \eta^\top_\mk \mc{T}_\s)$, where
\begin{subequations}\label{eq:BDG_matrices}
\begin{align}
\mc{H}^\w_\k &= \tau^z \otimes h^\w_\k \label{eq:H_wire} \\
\mc{H}^\s_\k &= \tau^z \otimes h^\s_\k + \tau^x \otimes \Delta_\k \mc{T}_\s, \label{eq:H_sc} \\
V_\k &= \tau^z \otimes t_\k, \label{eq:T_w_sc}
\end{align}
\end{subequations}
and where $\{\tau^\alpha\}_{\alpha=x,y,z}$ are Pauli matrices in particle-hole space. In writing Eqs.~(\ref{eq:H_BdG}, \ref{eq:BDG_matrices}), we have used the relations given in Eq.~(\ref{eq:TR_on_matrices}). The Green's function of the wire, $\mc{G}^\w_\k(\omega)$, is obtained by integrating out the SC,
\begin{subequations}\label{eq:Green_function}
\begin{align}
& \mc{G}^\w_\k(\omega) = [i\omega - \mc{H}^\w_\k - \Sigma_\k(\omega)]^{-1},\label{eq:G_w}  \\
& \Sigma_\k(\omega) = V^\dag_\k g^\s_\k(\omega) V_\k,\label{eq:self_E} \\
& g^\s_\k(\omega) = (i\omega - \mc{H}^\s_\k)^{-1},\label{eq:g_sc}
\end{align}
\end{subequations}
where $\Sigma_\k(\omega)$ is the self energy, and $g^\s_\k(\omega)$ is the Green's function of the parent SC in the absence of coupling to the wire.

Using Eqs.~\eqref{eq:H_sc} and~\eqref{eq:g_sc}, one can check that $g^\s_k(0)$ is Hermitian and obeys $\tau^y g^\s_\k(0) \tau^y= -g^\s_\k(0)$. It therefore has the following structure:
\begin{equation}\label{eq:g_sc_struct}
g^\s_\k(0) = \tau^z \otimes g^\tx{N}_\k + \tau^x \otimes g^\tx{A}_\k,
\end{equation}
where $g^\tx{N}_\k$ and $g^\tx{A}_\k$ are Hermitian matrices. This also means that the zero-frequency self energy has the same structure, $\Sigma_\k(0) = \tau^z \otimes \Sigma^\tx{N}_\k + \tau^x \otimes \Sigma^\tx{A}_\k$, with $\Sigma^\tx{N}_\k=t^\dag_\k g^\tx{N}_\k t_k$ and $\Sigma^\tx{A}_\k=-t^\dag_\k g^\tx{A}_\k t_k$ being the normal and anomalous parts, respectively. Upon rotating $g^\s_\k(0)$ in Eq.~\eqref{eq:g_sc} by the unitary transformation $\exp(i\pi\tau^x/4)$, and using Eqs.~\eqref{eq:H_sc} and~\eqref{eq:g_sc_struct}, it follows that
\begin{equation}\label{eq:Sigma_times_H}
(\Delta_\k \mc{T}_\s - ih^\s_\k)(g^\tx{A}_k + ig^\tx{N}_\k)=-\mathbbm{1}.
\end{equation}
One then arrives at
\begin{equation}\label{eq:Sigma_A_PSD}
\begin{split}
&\langle u | \Sigma^\tx{A}_\k | u \rangle = -\langle u | t^\dag_\k g^\tx{A}_\k t_\k | u \rangle =  -\Real \langle u | t^\dag_\k(g^\tx{A}_\k - i g^\tx{N}_\k) t_\k | u \rangle \\
&= \Real \langle u | t^\dag_\k(g^\tx{A}_\k - i g^\tx{N}_\k) (\Delta_\k \mc{T}_\s  - ih^\s_\k) (g^\tx{A}_\k + i g^\tx{N}_\k) t_\k| u \rangle \\
&= \Real \langle v |\Delta_\k \mc{T}_\s  - ih^\s_\k| v \rangle = \langle v |\Delta_\k \mc{T}_\s| v \rangle \ge 0,
\end{split}
\end{equation}
where $\vert u \rangle$ is an arbitrary vector, $| v \rangle \equiv (g^\tx{A}_\k + i g^\tx{N}_\k) t_\k| u \rangle$, and we have used the fact that $g^\tx{N}_\k$, $g^\tx{A}_\k$, and $h^\s_\k$ are Hermitian. Namely, we have proved that $\Sigma^\tx{A}_\k$ is PSD.

\emph{The topological invariant.---}We now follow Ref.~\cite{Haim2014time} and construct the $\mathbb{Z}_2$ topological invariant for a general gapped quasi 1d system in class DIII~\cite{OtherInvariantFormulations}. We then apply it to the system under consideration and show that, due to the positivity of $\Sigma_\k^{\mathrm{A}}$, the invariant always assumes its trivial value. We define the effective Hamiltonian of the wire system using its Green's function, $\mc{H}^\tx{eff}_\k = -[\mc{G}^\w_\k(0)]^{-1}$. By setting $\omega=0$ in Eq.~\eqref{eq:G_w} one obtains
\begin{equation}\label{eq:H_eff}
\mc{H}^\tx{eff}_\k = \tau^z\otimes(h^\w_\k + \Sigma^{N}_\k) + \tau^x\otimes\Sigma^\tx{A}_\k,
\end{equation}
where we have used the structure of $\Sigma_k(0)$ as given below Eq.~\eqref{eq:g_sc_struct}. This Hamiltonian obeys a time-reversal symmetry, $\mc{T}^\dag_\w \mc{H}^\tx{eff\ast}_\mk \mc{T}_\w=\mc{H}^\tx{eff}_\k$, as well as a chiral symmetry, $\tau^y \mc{H}^\tx{eff}_\k \tau^y=-\mc{H}^\tx{eff}_\k$, and is therefore in class DIII.

Written in the basis which diagonalizes the chiral symmetry, the Hamiltonian takes the form
\begin{equation}\label{eq:H_eff_rot}
e^{i\frac{\pi}{4}\tau^x} \mc{H}^\tx{eff}_\k e^{-i\frac{\pi}{4}\tau^x} = \begin{pmatrix} 0 & Q_\k \\ Q_\k^\dag & 0 \end{pmatrix}.
\end{equation}
We use the singular value decomposition to write $Q_k=U^\dag_\k D_\k V_\k$, where $U_\k$, $V_\k$ are unitary matrices and $D_\k$ is a square diagonal matrix with non-negative elements on its diagonal. By squaring $\mc{H}_\k^\tx{eff}$ it becomes apparent that the elements of $D_\k$ are the positive eigenvalues of $\mc{H}_\k^\tx{eff}$. Since $\mc{H}_\k^\tx{eff}$ is gapped, there are no zero elements on the diagonal of $D_\k$, and it is thus positive definite.

We can adiabatically deform $D_\k$ to the identity matrix without closing the gap, and therefore without changing the topological invariant. This in turn deforms the Hamiltonian, $\mc{H}^\tx{eff}_\k \rightarrow \tilde{\mc{H}}^\tx{eff}_\k$ , such that $\tilde{\mc{H}}^\tx{eff}_\k$ has two flat bands at energies $\pm 1$ (in the appropriate units), but the same eigenstates as $\mc{H}^\tx{eff}_\k$ (and therefore the same symmetries). $\tilde{\mc{H}}^\tx{eff}_\k$ is given by Eq.~\eqref{eq:H_eff_rot} with $Q_\k$ replaced by $\tilde{Q}_\k = U_\k^\dag V_\k$, which is now a unitary matrix. The TRS of $\tilde{\mc{H}}^\tx{eff}_\k$ implies that $\mc{T}^\dag_\w \tilde{Q}^\ast_\mk \mc{T}_\w = \tilde{Q}^\dag_\k$. Together with the unitarity of $\tilde{Q}_\k$ this dictates that for every eigenstate
\begin{equation}\label{eq:Q_es}
\tilde Q_\k | \alpha_{n,\k} \rangle = e^{i\theta_{n,\k}} | \alpha_{n,\k} \rangle,
\end{equation}
there is another eigenstate of $\tilde{Q}_k$, $\mc{T}^\dag_\w | \alpha_{n,\mk} \rangle^\ast$ with an eigenvalue $\exp(i\theta_{n,\mk})$. Thus, at the time-reversal invariant momenta, $k=0,\pi$, the eigenvalues of $\tilde{Q}_k$ come in Kramers' degenerate pairs.

\begin{figure}
\begin{tabular}{cc}
\hskip 0mm
\includegraphics[clip=true,trim =0mm 0mm 0mm 0mm,width=0.23\textwidth]{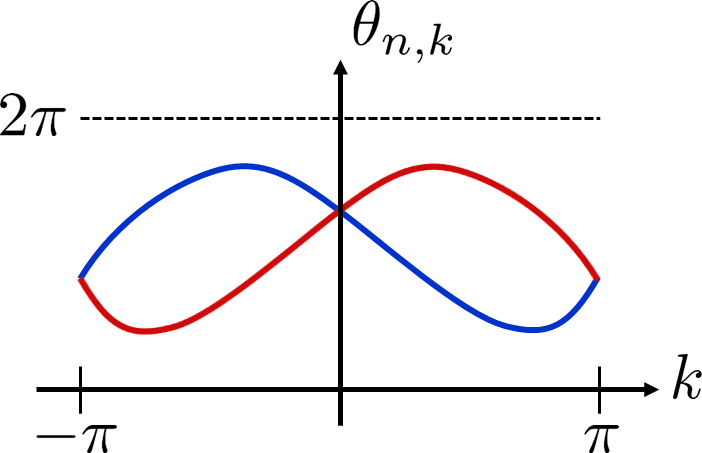}
\llap{\parbox[c]{8cm}{\vspace{-4.75cm}\footnotesize{(a)}}} &
\hskip -1mm
\includegraphics[clip=true,trim =0mm 28.3mm 0mm 0mm,width=0.23\textwidth]{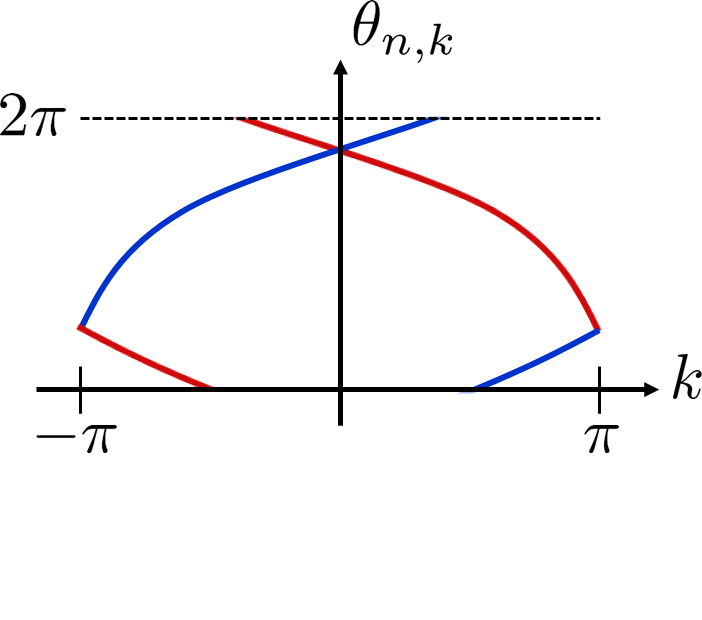} \llap{\parbox[c]{8cm}{\vspace{-4.75cm}\footnotesize{(b)}}}
\end{tabular}
\caption{Examples of spectra of the unitary matrix $\tilde{Q}_\k$ [see Eq.~\eqref{eq:H_eff_rot} and below], corresponding to (a) a topologically trivial case, and (b) a topologically nontrivial case. The eigenvalues of $\tilde{Q}_\k$ are phases given by $\{\exp(i\theta_{n,\k})\}_n$. Due to time-reversal symmetry the eigenvalues come in pairs, $\theta_{n,\k}$ and $\theta_{n,\mk}$, corresponding to the blue and red lines, respectively. The parity of the winding number of the blue (or red) line gives the class-DIII topological invariant in 1d. For a noninteracting system in proximity to an $s$-wave superconductor, the winding number of any angle $\theta_{n,\k}$ will always be zero [see Eq.~\eqref{eq:zero_winding} and below], rendering such a system topologically trivial.}\label{fig:theta_spectra}
\end{figure}

Considering the spectrum of $\tilde{Q}_\k$ as a function of $k\in[-\pi,\pi]$, it follows that the number of pairs of degenerate states at a given value $\theta$ cannot change by an odd number during an adiabatic change which leaves the gap of $\mc{H}^\tx{eff}_k$ open. The parity of the number of degenerate pairs is therefore a topological invariant. Alternatively stated, upon dividing the eigenvalues of $\tilde{Q}_\k$ to two groups $\{ \exp(i\theta_{n,\k}^\tx{I}) \}_n$ and $\{ \exp(i\theta_{n,\k}^\tx{II}) \}_n$, related by time reversal, $\theta_{n,\k}^\tx{II}=\theta^\tx{I}_{n,\mk}$, the topological invariant is given by
\begin{equation}\label{eq:Winding_parity}
\nu_\tx{1d} = (-1)^W \hskip 5mm ;\hskip 5mm
W = \sum_n \frac{1}{2\pi} \int_{k=-\pi}^{k=\pi} \tx{d}\theta_{n,\k}^\tx{I},
\end{equation}
namely, the parity of the sum of windings of $\{\theta_{n,\k}^\tx{I}\}_n$. Figure~\ref{fig:theta_spectra} presents examples of trivial and topological spectra of $\tilde{Q}_\k$.

Eq.~\eqref{eq:Winding_parity} is correct for any quasi 1d system in class DIII. Let us now concentrate on the system at hand, namely one which is given by Eq.~\eqref{eq:H}, with $\Delta_\k \mc{T}_\s$ being PSD. Inserting Eq.~\eqref{eq:H_eff} in Eq.~\eqref{eq:H_eff_rot}, one arrives at $Q_\k =\Sigma^\tx{A}_\k - i (h^\w_\k+\Sigma^\tx{N}_\k)$. From the positivity of $\Sigma^\tx{A}_\k$, derived in Eq.~\eqref{eq:Sigma_A_PSD}, and the fact that $h^\w_\k$ and $\Sigma^\tx{N}_\k$ are Hermitian, it follows that
\begin{equation}\label{eq:zero_winding}
\begin{split}
0 \le & \langle \alpha_{n,\k} | \Sigma^\tx{A}_\k | \alpha_{n,\k} \rangle
= \Real \langle \alpha_{n,\k} | Q_\k | \alpha_{n,\k} \rangle  \\
= & \Real \langle \alpha_{n,\k} | U^\dag_\k D_\k V_\k  | \alpha_{n,\k} \rangle  \\
= &\Real \langle \alpha_{n,\k} | \tilde{Q}_\k V^\dag_\k D_\k V_\k  | \alpha_{n,\k} \rangle  \\
= & 2\cos\theta_{n,k} \cdot
\langle V_\k \alpha_{n,\k} | D_\k | V_\k \alpha_{n,\k} \rangle,
\end{split}
\end{equation}
and since $D_\k$ is positive definite, we conclude that $\cos\theta_{n,k} \ge 0$ for all $n$ and $k$. Namely none of the phases $\theta_{n,k}$ can have a non-zero winding number as $k$ changes from $-\pi$ to $\pi$, which in particular means that the topological invariant, Eq.~(\ref{eq:Winding_parity}), is always trivial, $\nu_\tx{1d}=1$.

\emph{Two dimensions.---}We wish to generalize our result to the case of a 2d system in proximity to a bulk superconductor, as depicted in Fig.\hyperref[fig:setup]{~\ref{fig:setup}(b)}. The combined system is described by the Hamiltonian of Eq.~\eqref{eq:H}, with $k\rightarrow \bs{k}=(k_x,k_y)$. All the results, excluding Eq.~\eqref{eq:Winding_parity}, are still valid in the 2d case under this substitution. The $\mathbb{Z}_2$ two-dimensional topological invariant can be obtained from the 1d invariant by~\cite{NonOrthorhombicSystems}
\begin{equation}\label{eq:2d_top_inv}
\nu_\tx{2d} = \nu_\tx{1d}[\mc{H}^\tx{eff}_{k_x=0,k_y}]\nu_\tx{1d}[\mc{H}^\tx{eff}_{k_x=\pi,k_y}].
\end{equation}
Before proving Eq.~\eqref{eq:2d_top_inv}, let us first draw from it our main conclusion. The Hamiltonians $\mc{H}^\tx{eff}_{k_x=0,k_y}$ and $\mc{H}^\tx{eff}_{k_x=\pi,k_y}$ both belong to class DIII in 1d, and are of the form of Eq.~\eqref{eq:H_eff} with a PSD anomalous part. Consequently, as we proved above, both $\mc{H}^\tx{eff}_{k_x=0,k_y}$ and $\mc{H}^\tx{eff}_{k_x=\pi,k_y}$ are topologically trivial. From Eq.~\eqref{eq:2d_top_inv} it then follows that the 2d Hamiltonian $\mc{H}^\tx{eff}_{k_x,k_y}$ is trivial as well~\cite{WeakInv}.

We now argue that the two-dimensional topological invariant in class DIII is given by Eq.~\eqref{eq:2d_top_inv}. This is most readily seen by considering a semi-infinite system with periodic boundary conditions in the $x$ direction, and an edge along the line $y=0$.
The non-trivial phase is characterized by having an odd number of helical edge modes. At the edge of such system, at every energy inside the bulk gap, there must be an odd number of Kramers' pairs of edge states, similarly to the case of the two-dimensional topological insulator~\cite{Kane2005quantum,Fu2006}. Let us focus on $E=0$ (which is in the middle of the gap, due to particle-hole symmetry).
At $k_x=0$, the number of Kramers' pairs is equal to the $\mathbb{Z}_2$ invariant of the corresponding DIII one-dimensional Hamiltonian $\mathcal{H}^{\mathrm{eff}}_{k_x=0, k_y}$.  The same is true at the other time reversal invariant momentum, $k_x=\pi$. Due to time reversal and chiral symmetries, the number of zero energy Kramers' pairs at momenta away from $k_x=0,\pi$ must be even. Therefore, the parity of the total number of Kramers' pairs at $E=0$ is equal to $\nu_\tx{1d}[\mc{H}^\tx{eff}_{k_x=0,k_y}]\nu_\tx{1d}[\mc{H}^\tx{eff}_{k_x=\pi,k_y}]$, which is the right hand side of Eq.~\eqref{eq:2d_top_inv}.

{\emph{Extension to non-translationally invariant systems.---}}
So far, we assumed that the system is translationally invariant along the direction of the wire in the 1d case, or in the plane of the system in the 2d case [Figs.~\hyperref[fig:setup]{\ref{fig:setup}(a)} and~\hyperref[fig:setup]{\ref{fig:setup}(b)}, respectively]. However, our results holds even without translational symmetry, e.g., in the presence of disorder.

To see this, consider a disordered system in either 1d or 2d, coupled to a superconductor. Imagine a disorder realization which is periodic in space, with a period that is much larger than any microscopic length scale (in particular, the induced superconducting coherence length).
By the arguments presented in the preceding sections, the resulting translationally invariant system is topologically trivial. Hence, at its boundary there are no topologically non-trivial edge states. Since the size of the unit cell is much larger than the coherence length, the periodicity of the system cannot matter for the existence or the lack of edge states. Therefore, a \emph{single} unit cell corresponds to a finite disordered system, which (as its size tends to infinity) is in the topologically trivial phase, as well.

\emph{Discussion.---}We have examined a general time-reversal symmetric system of noninteracting electrons in proximity to a bulk conventional superconductor. It was shown that irrespective of any details of the electronic structure of the system and the form of its coupling to the superconductor, the system is always in a topologically trivial phase.

More generally, the condition for the system to be trivial is that the pairing matrix of the parent superconductor, $\Delta_\k$, satisfies that $\Delta_\k\mc{T}_\s$ is positive semi-definite, where $\mc{T}_\s$ is the representation of time-reversal [see Eqs.~\eqref{eq:H} and~\eqref{eq:TR}]. In particular, this condition applies for example to the case of a conventional $s$-wave superconductor, in which the gap function has the same sign on all bands (and there is no interband pairing). The parent superconductor can have any number of bands and an arbitrary form of spin-orbit coupling.

These results have implications for the search for realizations of time-reversal invariant topological superconductors in class DIII.
In order to avoid the trivial fate of the system, one has to either invoke strong enough electron-electron interactions~\cite{Gaidamauskas2014majorana,Haim2014time,Klinovaja2014Kramers,Klinovaja2014time,Danon2015interaction,Pedder2015,Haim2016interaction}, or use a parent SC for which $\Delta_\k\mc{T}_\s$ is \emph{not} positive semi-definite. This can be an unconventional SC~\cite{Wong2012majorana,Zhang2013time,Nakosai2013majorana}, or a combination of two SCs in a $\pi$ junction~\cite{Dahlhaus2010,Keselman2013inducing,Schrade2015proximity}.

\emph{Acknowledgements.---}We have benefited from discussions with I. C. Fulga and Y. Schattner. E. B. was supported by the Minerva foundation, by a Marie Curie Career Integration Grant (CIG), and by the European Research Council (ERC) under the European Union's Horizon 2020 research and innovation programme (grant agreement No. 639172). Y. O. was supported by the Israeli Science Foundation (ISF), by the Minerva foundation, by the Binational Science Foundation (BSF) and by the ERC, grant No. 340210 (FP7/2007-2013). K. F. was supported by the Danish National Research Foundation and by the Danish Council for Independent Research $|$ Natural Sciences.

\bibliography{Refs_impossibility_of_using_s_wave}

\begin{thebibliography}{57}
\bibitem [{Rev()}]{ReverseProximityEffect}%
  \BibitemOpen
  \href@noop {} {}\bibinfo {note} {The reverse proximity effect in the SC may
  reduce $\Delta_k$ near the interface with the wire, compared to the bulk
  value. This effect should be small when the density of states in the SC
  (within a coherence length away from the interface) is large compared to the
  density of states in the wire, or alternatively, when the coupling to the
  wire is weak.}\BibitemShut {Stop}%
\bibitem [{Cmp()}]{CmplxNumTimesPSD}%
  \BibitemOpen
  \href@noop {} {}\bibinfo {note} {More generally, $\Delta_\k\mc{T}_\s$ can be
  a PSD matrix times some complex number that does not depend on $k$. The phase
  of this complex number can always be absorbed in the definition of
  $\mc{T}_\s$, rendering $\Delta_\k\mc{T}_\s$ PSD.}\BibitemShut {Stop}%
\bibitem [{SOC()}]{SOCinSC}%
  \BibitemOpen
  \href@noop {} {}\bibinfo {note} {Notice this does not exclude the possibility
  that the superconductor has spin-orbit coupling.}\BibitemShut {Stop}%
\bibitem [{Non()}]{NonOrthorhombicSystems}%
  \BibitemOpen
  \href@noop {} {}\bibinfo {note} {Note that most generally $k_x$ and $k_y$
  should be considered as coordinates along the reciprocal primitive vectors,
  $\bs{G}_{1,2}$, in units of $|\bs{G}_{1,2}|/2\pi$, respectively.}\BibitemShut
  {Stop}%
\bibitem [{Wea()}]{WeakInv}%
  \BibitemOpen
  \href@noop {} {}\bibinfo {note} {We note that since \emph{both}
  $\mc{H}^\tx{eff}_{k_x=0,k_y}$ and $\mc{H}^\tx{eff}_{k_x=\pi,k_y}$ (and
  similarly $\mc{H}^\tx{eff}_{k_x,k_y=0}$ and $\mc{H}^\tx{eff}_{k_x,k_y=\pi}$)
  are trivial, the weak topological indices are trivial as well.}\BibitemShut
  {Stop}%
\end{thebibliography}
\end{document}